\documentclass[twocolumn,trackchanges]{aastex701}

\begin{document}

\title{Mass-Metallicity Relation of the Fossil Disk in Six Nearby Galaxies using JWST RGB Stars}

\author[orcid=0000-0002-7507-7227]{Avinash Ck}
\affiliation{Instituto Nacional de Astrofísica, Óptica y Electrónica, Luis Enrique Erro 1, Tonantzintla 72840, Puebla, México}
\email[show]{avinash@inaoe.mx}  

\author[orcid=0000-0002-4677-0516]{Divakara Mayya} 
\affiliation{Instituto Nacional de Astrofísica, Óptica y Electrónica, Luis Enrique Erro 1, Tonantzintla 72840, Puebla, México}
\email{dmayya@gmail.com}

\author[orcid=0000-0002-7922-8440]{Alessandro  Bressan}
\affiliation{SISSA, Via Bonomea 265, I-34136 Trieste, Italy}
\affiliation{Osservatorio Astronomico di Padova—INAF, Vicolo dell’Osservatorio 5, I-35122 Padova, Italy}
\affiliation{Purple Mountain Observatory, Chinese Academy of Sciences, Nanjing 210023, People’s Republic of China}
\email{sbressan@sissa.it}

\author[orcid=0000-0002-8351-8854]{Jairo Andres Alzate Trujillo}
\affiliation{Instituto de Astrofísica de Andalucía – Consejo Superior de Investigaciones Científicas (IAA-CSIC), 
S/N,E-18008, Granada, Spain}
\email{jalzate@cefca.es}

\author[orcid=0000-0002-6301-3269]{Léo Girardi}
\affiliation{Osservatorio Astronomico di Padova—INAF, Vicolo dell’Osservatorio 5, I-35122 Padova, Italy}
\email{leo.girardi@inaf.it}

\author[orcid=0009-0004-7083-8578]{Shaunak Mishra}
\affiliation{LIRA, Observatoire de Paris, Université PSL, Sorbonne Université, Université Paris Cité, CNRS, 92190 Meudon, France}
\email{Shaunak.MISHRA@obspm.fr}

\author[orcid=0000-0002-1046-1500]{Bolivia Cuevas-Otahola}
\affiliation{Instituto de Astronomía, Universidad Nacional Autónoma de México, Box 70-264, México City, México}
\email{b.cuevas.otahola@gmail.com}

\begin{abstract}
We here present the first ever mass-metallicity relation (MZR) of fossil disk populations using the resolved stellar populations in the James Webb Space Telescope Near-Infrared Camera (JWST/NIRCam) images of six nearby galaxies. Both the mass and metallicity were determined using the Red Giant Branch (RGB) stars, employing methods that have been well established to study the halo populations of nearby galaxies using the Hubble Space Telescope images. Specifically, we obtained the mass and metallicity by comparing the color, magnitude and number density near the tip of the RGB (TRGB) stars in the F115W$-$F200W vs F200W color-magnitude diagrams (CMDs) with that of the PARSEC isochrones of 4 and 13~Gyr. The RGB metallicities of the disks of nearby galaxies closely match those found in galaxies at redshifts $z>8$, with the metallicity range following a power-law relation with the fossil disk mass. The power-law index of the fossil disk MZR, $\gamma\approx0.26$, compares well with the typical value found in MZRs obtained using gas metallicities across redshifts.
Through this work, we illustrate the utility of JWST/NIRCam images for obtaining the mass and metallicity of the fossil disks of nearby galaxies at precisions that were not possible earlier, which opens up a new method to carry out the near-field cosmological studies.
\end{abstract}



\section{Introduction}
The mass-metallicity relation (MZR), which is the correlation between a galaxy's stellar mass and metallicity \citep{mzr_og1, mzr_og2}, is one of the most fundamental scaling relations in galaxy evolution \citep[][and references therein]{fmr}, where galaxies with larger stellar mass have higher metallicities. This relation is sculpted by the interplay between gas-accretion, star formation and feedback-driven outflows \citep{Edmunds1990, LehnertHeckman1996, Garnett2002, mzr_sdss, KobayashiSpringelWhite2007,Scannapieco2008,Spitoni2010}.

\begin{deluxetable}{ccccc|cccccccccc}
\tablecaption{Observed and derived properties of galaxies in our sample \label{tab:prop}}
\tabletypesize{\scriptsize}
\tablehead{\colhead{} & \colhead{} & \colhead{} & \colhead{} & \colhead{} & \colhead{} &
\colhead{} & \multicolumn{2}{c}{$\log({\rm Z}/{\rm Z}_{\sun})$} & \colhead{}  & \multicolumn{2}{c}{$\log($M$_*$/M$_{\sun})$} & \colhead{} & \colhead{} & \colhead{}\\
\colhead{ID} & \colhead{Morph.} & \colhead{$i$} & \colhead{P.A.} & \colhead{DM}  &
\colhead{N$_{\rm grid}$} & \colhead{DM} & \colhead{4~Gyr} & \colhead{13~Gyr} & \colhead{A$_V$}  & \colhead{4~Gyr} & \colhead{13~Gyr}  & \colhead{$\Delta$M$_{\rm corr}$} & \colhead{M$_{\rm frac}$} & \colhead{Ref.}\\
\colhead{} & \colhead{} & \colhead{deg} & \colhead{deg} & \colhead{mag}  & \colhead{} & \colhead{mag} & \colhead{} & \colhead{} & \colhead{mag} &
\colhead{} & \colhead{} & \colhead{dex}& \colhead{} & \colhead{}}
\colnumbers
\startdata
NGC\,628$^{\dagger}$ & SA(s)c  & 9  & 21    & 29.96  & 79 & 29.81 & $-$0.73 & $-$0.98 & 0.14 & 9.85$\pm$0.06  & 10.19$\pm$0.05  & 0.12 & 0.16 &1,6\\
NGC\,5194             & SA(s)bc & 20 & 12   & 29.67 & 44 & 29.40 & $-$0.63 & $-$0.89 & 0.27 & 10.29$\pm$0.06 & 10.63$\pm$0.07  & 0.13 & 0.18 & 2,7\\
NGC\,5236             & SAB(s)c & 20 & 225  & 28.41 & 35 & 28.26 & $-$0.62 & $-$0.87 & 0.22 & 10.11$\pm$0.07 & 10.46$\pm$0.08  & 0.11 & 0.31 & 3,8\\
NGC\,4449             & IBm     & 45 & 45   & 28.02 & 11 & 28.00 & $-$0.91 & $-$1.16 & 0.09 & 8.87$\pm$0.05  & 9.22$\pm$0.06   & 0.11 & 0.73& 4,9\\
NGC\,4490             & SB(s)d  & 65 & 115  & 29.70 & 12 & 29.67 & $-$1.13 & $-$1.40 & 0.27 & 9.18 $\pm$0.08  & 9.52$\pm$0.09   & 0.14  & 0.42 & 5,9\\
NGC\,4485             & IB(s)m  & 44 & 346 & 29.70 & 6 & 29.67 & $-$1.06 & $-$1.29 & 0.02 & 8.63$\pm$0.11  & 8.97$\pm$0.11   & 0.20  & 0.52 & 5,9\\
 \enddata
\tablecomments{$\dagger$~Galaxy analyzed in Paper~I. \textbf{Columns 2--5 (from literature):} (2)~Morphological type from NASA Extragalactic Database~(\href{https://ned.ipac.caltech.edu/}{NED}); (3--5)~inclination, position angle and literature distance modulus using HST TRGBs, taken from references in Column 15. \textbf{Columns 6--14 (this work):} (6)~Number of spatial bins used for TRGB analysis with 50\% completeness magnitude at least 1~mag below the TRGB; (7)~Distance modulus derived in this work; (8--9)~median metallicity from TRGB measurements using 4 and 13~Gyr isochrones, respectively; (10)~Interstellar Medium (ISM) extinction derived from TRGB following the method described in Paper~I; (11--12)~RGB population stellar masses for assumed ages of 4 and 13~Gyr after applying the correction from column~13; (13)~Correction in stellar mass due to contamination from younger stars in the TRGB window (14) Fraction of mass obtained directly from TRGB counts with respect to total stellar mass obtained from exponential fitting; (15)~References for inclination, position angle, and literature distance modulus. [1]~\citet{ngc628_disc_prop}, [2]~\citet{m51_disc_prop}, [3]~\cite{m83_disc_prop}, [4]~\citet{ngc4449_disc_prop}, [5]~\cite{coming}, [6]~\citet{trgb_ngc628_lit}, [7]~\citet{trgb_m51}, [8]~\citet{trgb_m83}, [9]~\citet{trgb_ngc4449}.}
\end{deluxetable}

At $z=0$, this relation is particularly well established through surveys such as the Sloan Digital Sky Survey \citep[SDSS;][]{sdss}. With the advent of efficient spectrographs at near-infrared (NIR) wavelengths, the MZR has been extended well beyond $z=0$ \citep[e.g.][]{high_z_mzr_1,high_z_mzr_2,high_z_mzr_3}. Studies indicate that its overall functional form remains broadly similar with cosmic time, while its normalization evolves. Most strikingly, recent James Webb Space Telescope (JWST) spectroscopy has pushed metallicity measurements out to $z\approx8$ and beyond, probing the MZR at the earliest observable epochs of galaxy formation. However, these high-redshift anchor points carry large uncertainties as they are generally derived from strong optical emission lines and spectral energy distribution (SED) modeling. Given the intrinsic faintness and complexity of high-redshift galaxies, the observational uncertainties that get carried forward into mass and metallicity estimates are quite significant. 

An alternative to high-redshift observation for studying early epochs of galaxy evolution is to study the fossil resolved stellar populations of nearby galaxies where the observational uncertainties are minimal. Nearby galaxies host old stellar populations that 
trace the current disk which was probably formed during the early epochs of galaxy assembly, analogous to a young disk at high redshift. Although the morphology and mass of the fossil disk could have changed  \citep[e.g.][]{Minchev}, the chemical properties are still preserved. This local, resolved approach is in the tradition of near-field cosmology, in which the present-day fossil record of nearby galaxies is used to constrain conditions and processes at earlier cosmic times.

One of the key tools enabling such studies is the relationship between red giant branch (RGB) star color and metallicity. The tip of the red giant branch (TRGB), the sharp luminosity-function discontinuity marking the onset of the helium flash in stars with initial masses $\lesssim$~2~M$_\odot$ and ages $>$~4~Gyr, has long served as a standard candle for distance measurement \citep{trgb1, trgb2, trgb2b, trgb3, trgb4, trgb5, trgb6, trgb7, trgb8}. Its color and magnitude also depend systematically on metallicity, a relation grounded in stellar evolution theory \citep{trgb_color1, trgb_color2} and calibrated empirically using the Galactic  globular clusters \citep{trgb_met1}. Since then, the TRGB color-metallicity relation has been used to derive metallicities in nearby galaxies where the RGBs can be identified in resolved stellar populations using the ground-based  observations in the Magellanic Clouds \citep{trgb_met_smc} and M31/33 \citep{trgb_met_m31} and Hubble Space Telescope (HST) observations of nearby dwarf and irregular galaxies \citep{trgb_met2, trgb_met3, trgb_met4}, firmly establishing the RGB/TRGB color as a robust metallicity tracer beyond its original distance-measurement role.


\begin{figure*}
    \centering
    \includegraphics[width=0.33\linewidth]{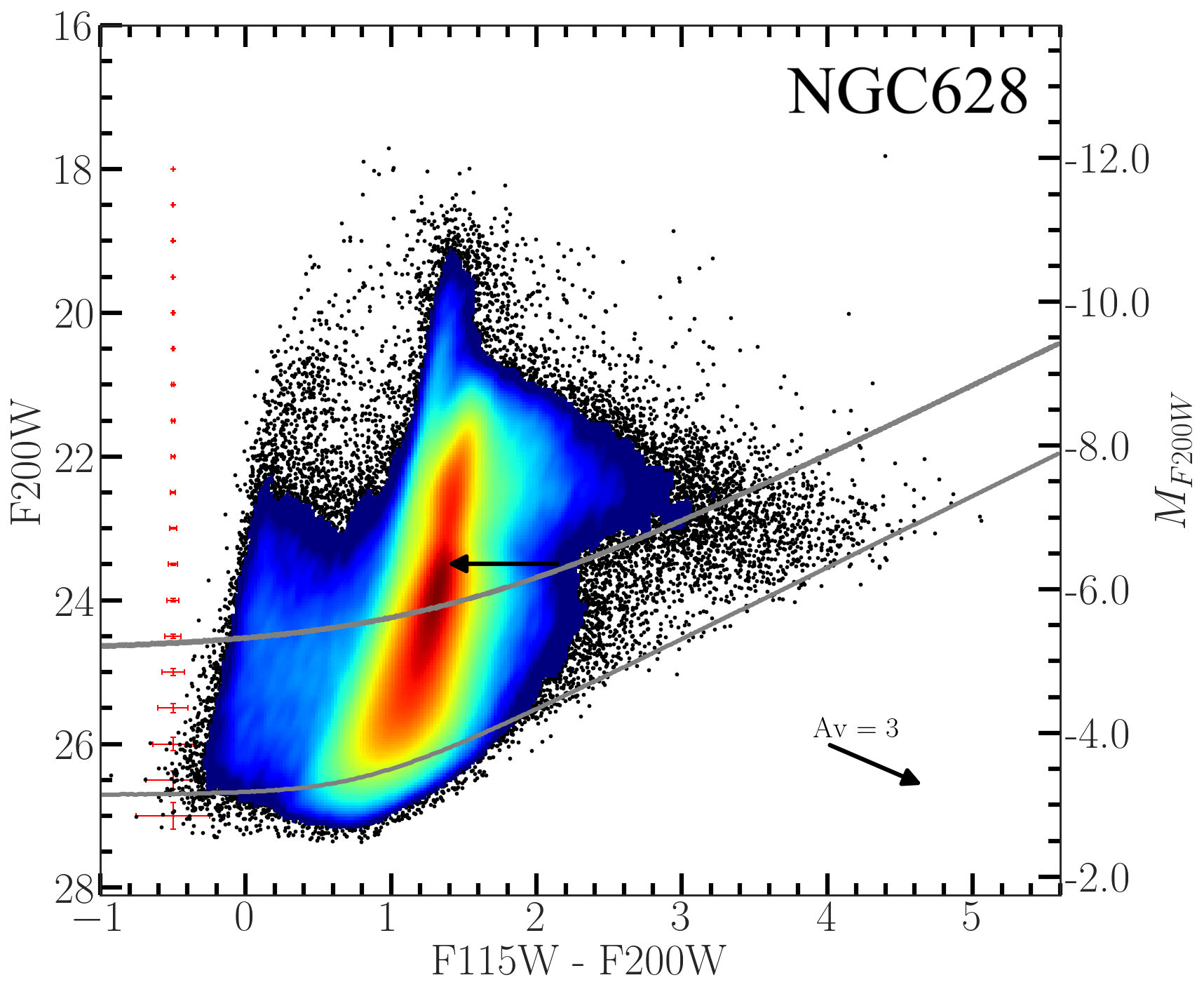}%
    \includegraphics[width=0.33\linewidth]{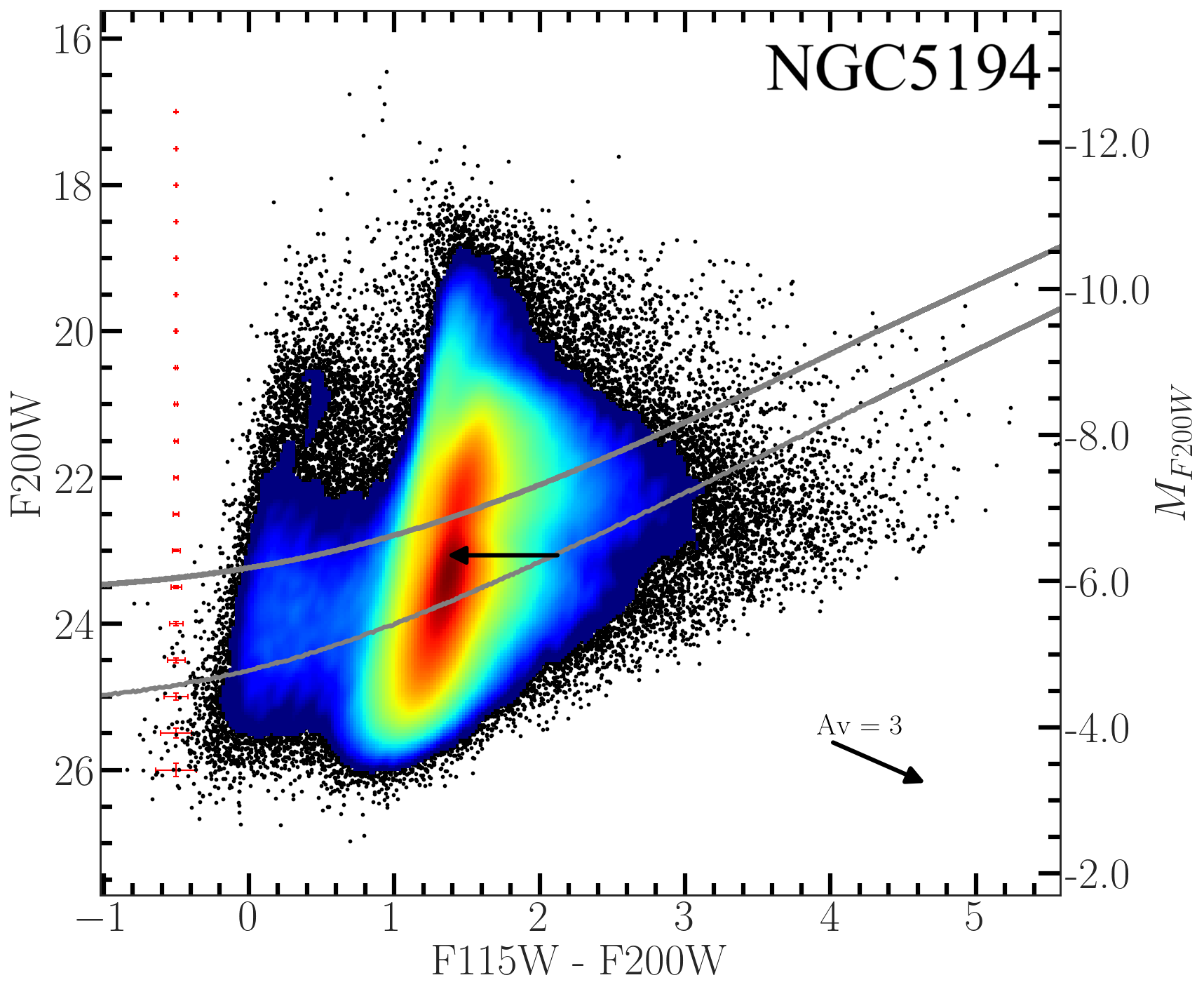}%
    \includegraphics[width=0.33\linewidth]{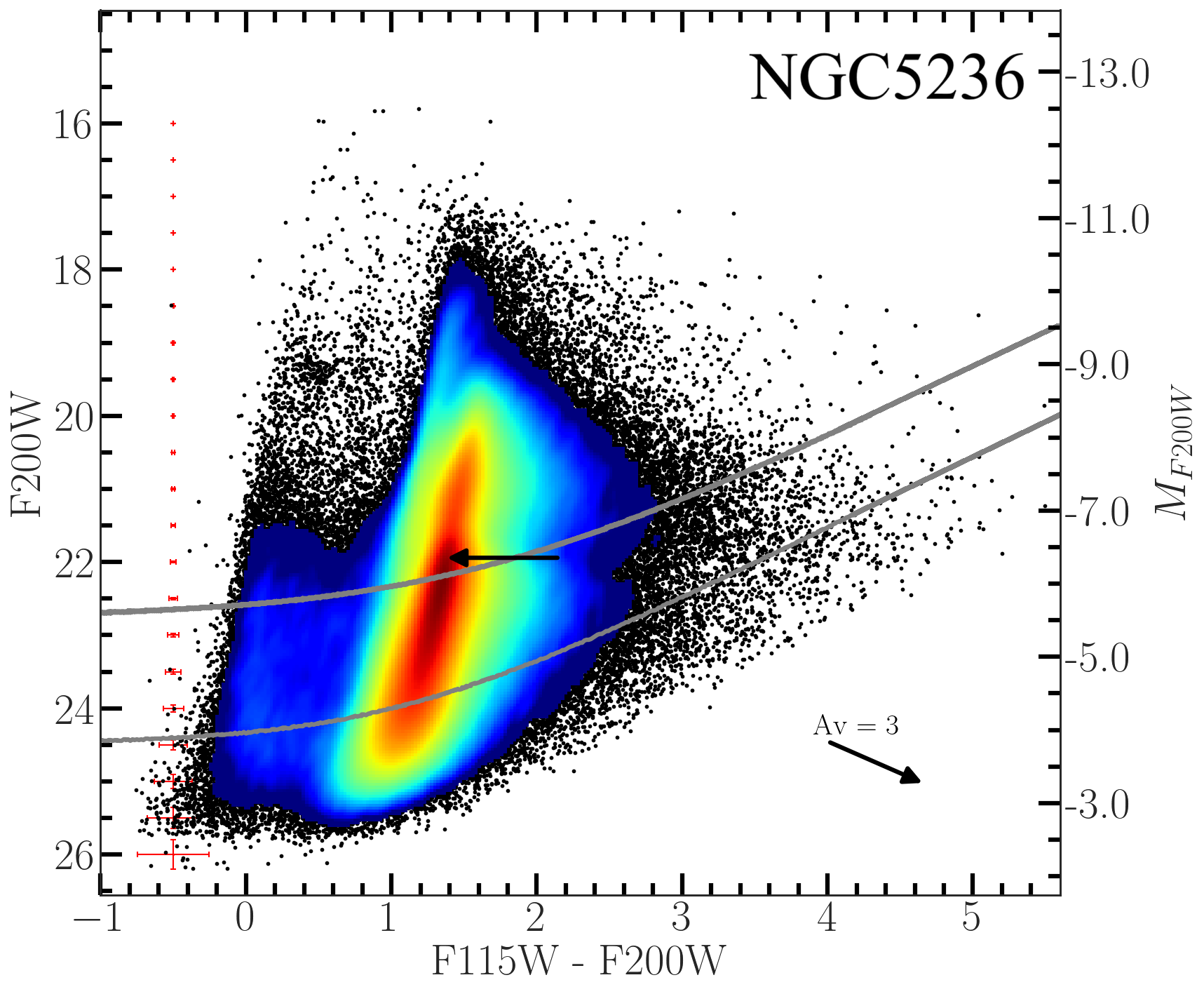}
    \includegraphics[width=0.33\linewidth]{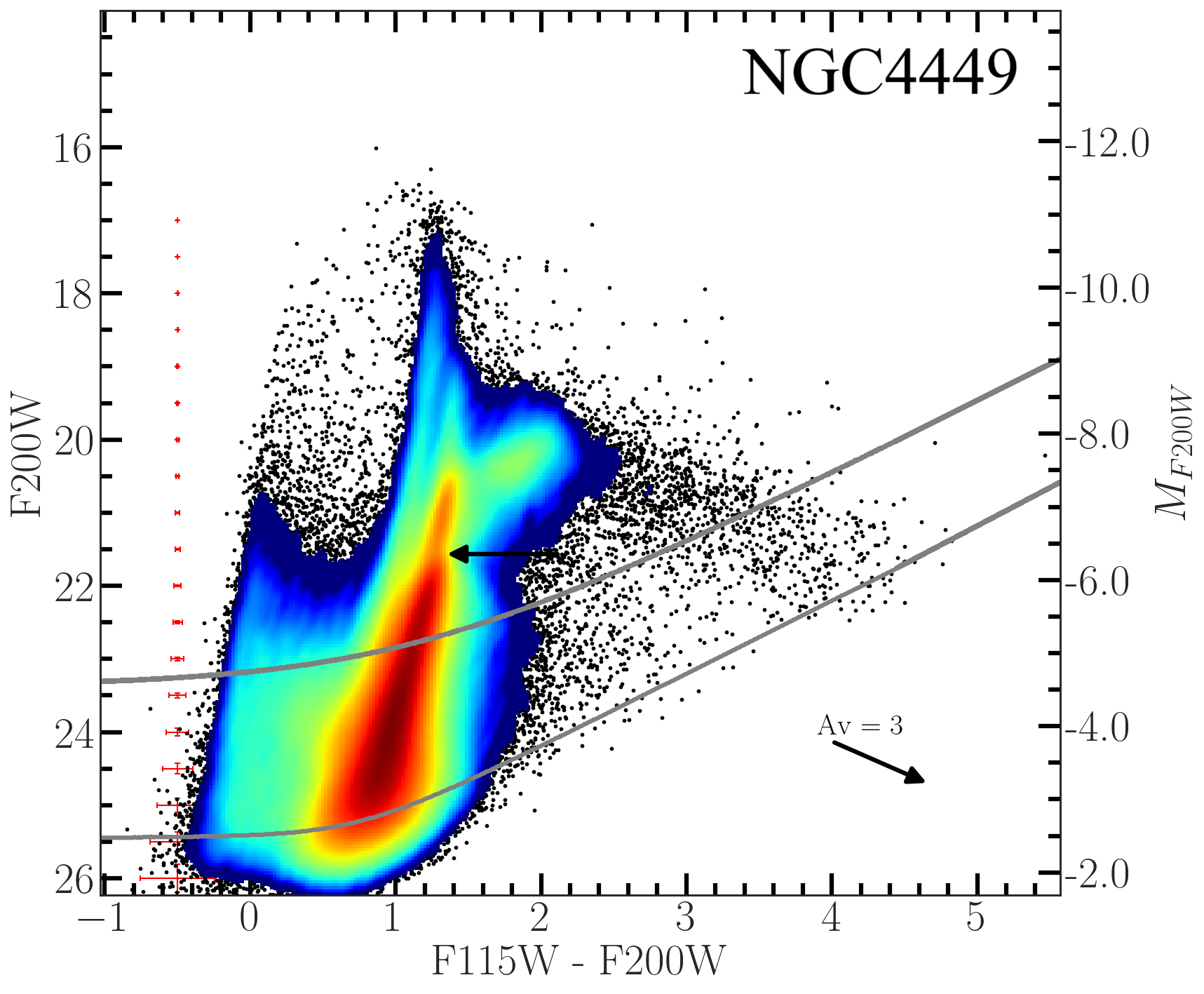}%
    \includegraphics[width=0.33\linewidth]{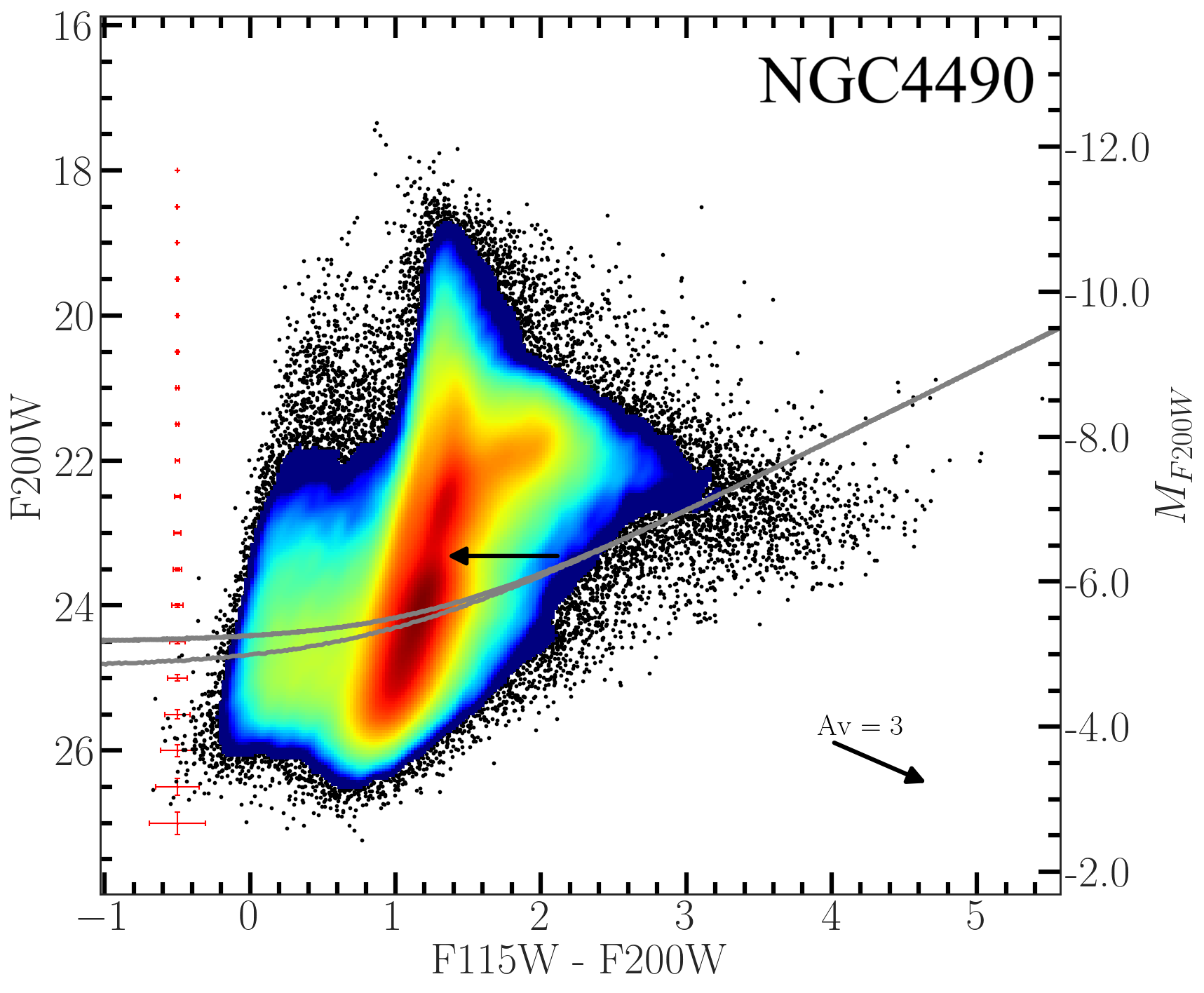}%
    \includegraphics[width=0.33\linewidth]{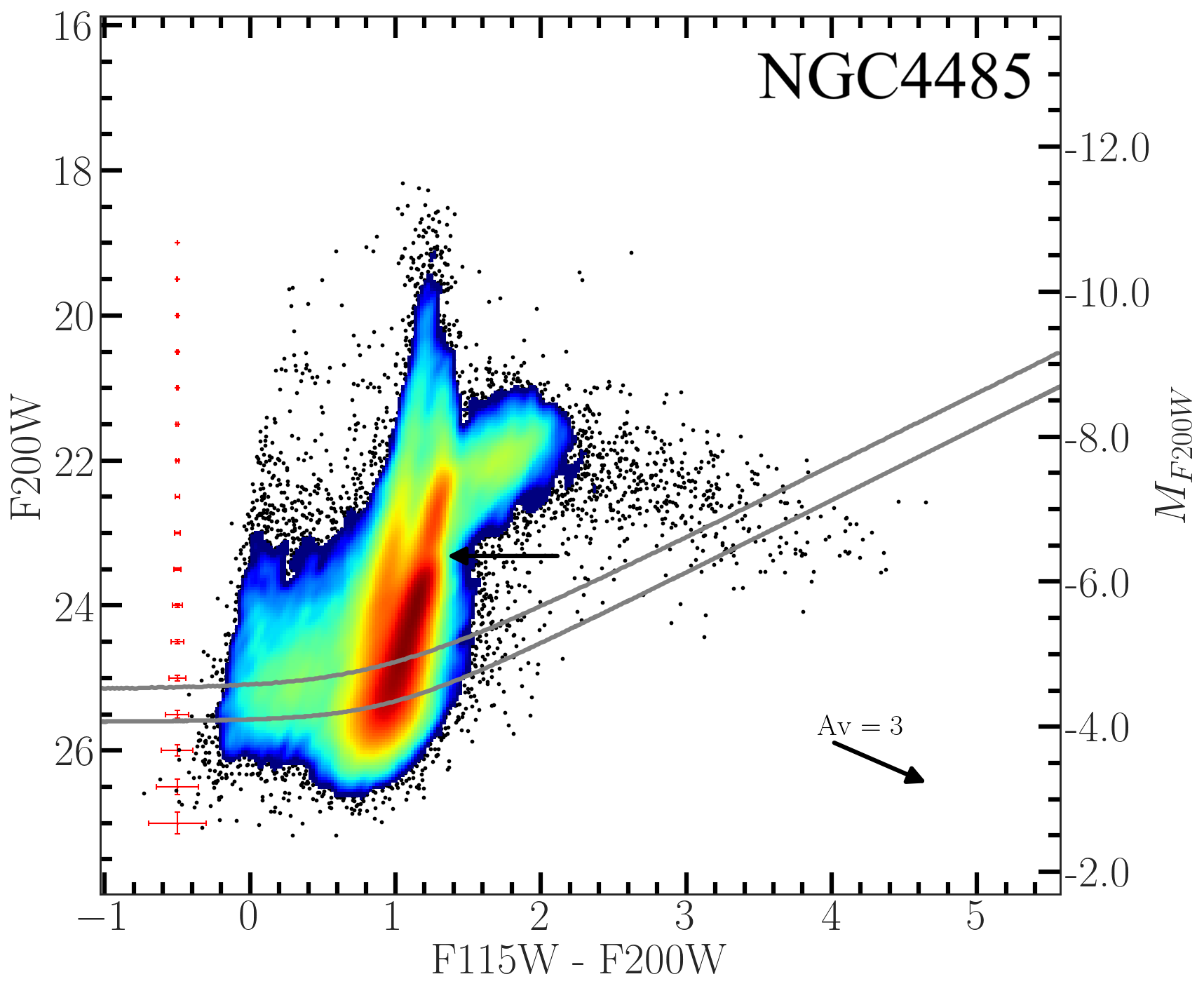}%
    \caption{Hess Diagram of F115W$-$F200W vs. F200W CMD  containing stars over the entire FoV. Black points indicate points above 90$^{\rm th}$~percentile of kernel density. \textit{ Top:} (left to right) NGC\,628, NGC\,5194 and NGC\,5236. \textit{ Bottom:} (left to right) NGC\,4449, NGC\,4490 and NGC\,4485. The typical errors on magnitude and color are given as a vertical sequence of red error bars at the left part of the diagrams. The 50\% completeness limit is shown by the gray curves, with the upper and lower curves corresponding to the inner and outer parts of the galaxies, respectively. The reddening vector corresponding to $A_V$=3~mag is shown at the bottom-right corner of each panel. A horizontal arrow is placed at the absolute magnitude M$_{F200W}=-6.354$ mag (right axis) corresponding to fixed age ($\log$(Age/yr)=10) and metallicity (Z=0.002 or equivalently $\log(Z/Z\sun)=-0.93$), and using the distance modulus from column~7 in Table~\ref{tab:prop}. A slight shift between the TRGB point for each galaxy and arrow location can be visualized which is due to the changes in the metallicity. }
    \label{fig:cmds}
\end{figure*}

\citet{Harmsen2017} and \citet{Bell2017} have taken advantage of the relatively low crowding and minimal extinction in the halos of galaxies to not only use the RGBs as metallicity tracers, but also use their number densities to estimate stellar masses of halos using the HST optical observations. This has  allowed them to construct the stellar halo MZR. However, extending these techniques to galaxy disks have been hampered by higher stellar density and variable dust extinction. With the advent of JWST/NIRCam, it is now possible to detect individual RGB stars in disks of nearby galaxies \citep[e.g.][]{Correnti_2025} and use the RGB colors for metallicity measurements \citep[][hereafter Paper I]{trgb_ngc628}.

In this Letter, we use the JWST data of six nearby galaxies spanning a broad range of morphology and stellar mass to address the mass-metallicity relation of the fossil disks of galaxies. In Section~2, we briefly describe the galaxy sample utilized in this work, along with the JWST data. In Section~3, we detail the technique utilized to obtain TRGB color-magnitude and stellar mass from RGB number density. In Section~4, we present the stellar metallicity obtained from TRGB color-magnitude measurements and the MZR of the fossil disks and compare it with MZR at $z>8$. Conclusions are given in Section~5.

\section{Data and Stellar Catalog}
\label{sec:data}

\begin{figure*}
    \centering
    \includegraphics[width=0.33\linewidth]{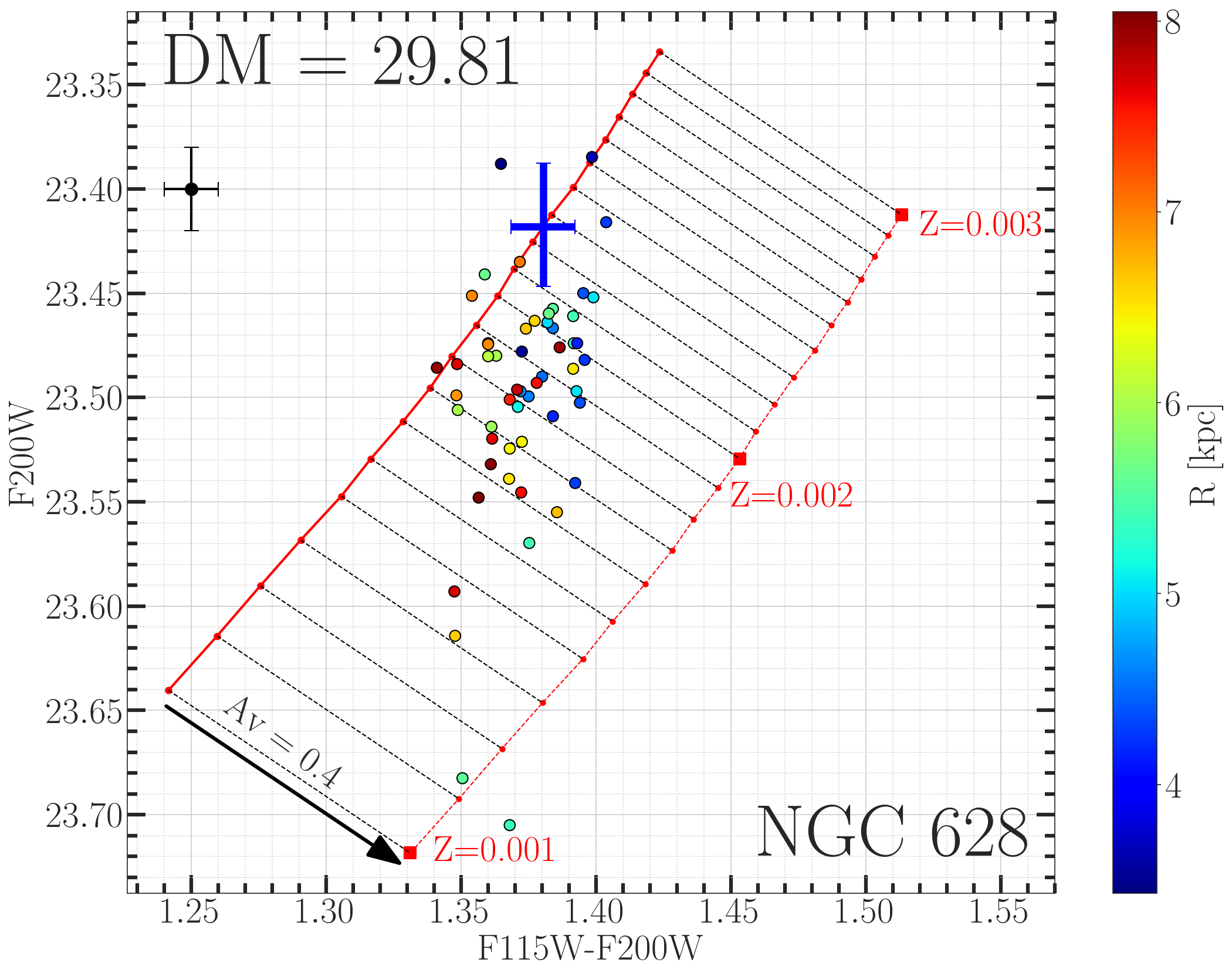}%
    \includegraphics[width=0.33\linewidth]{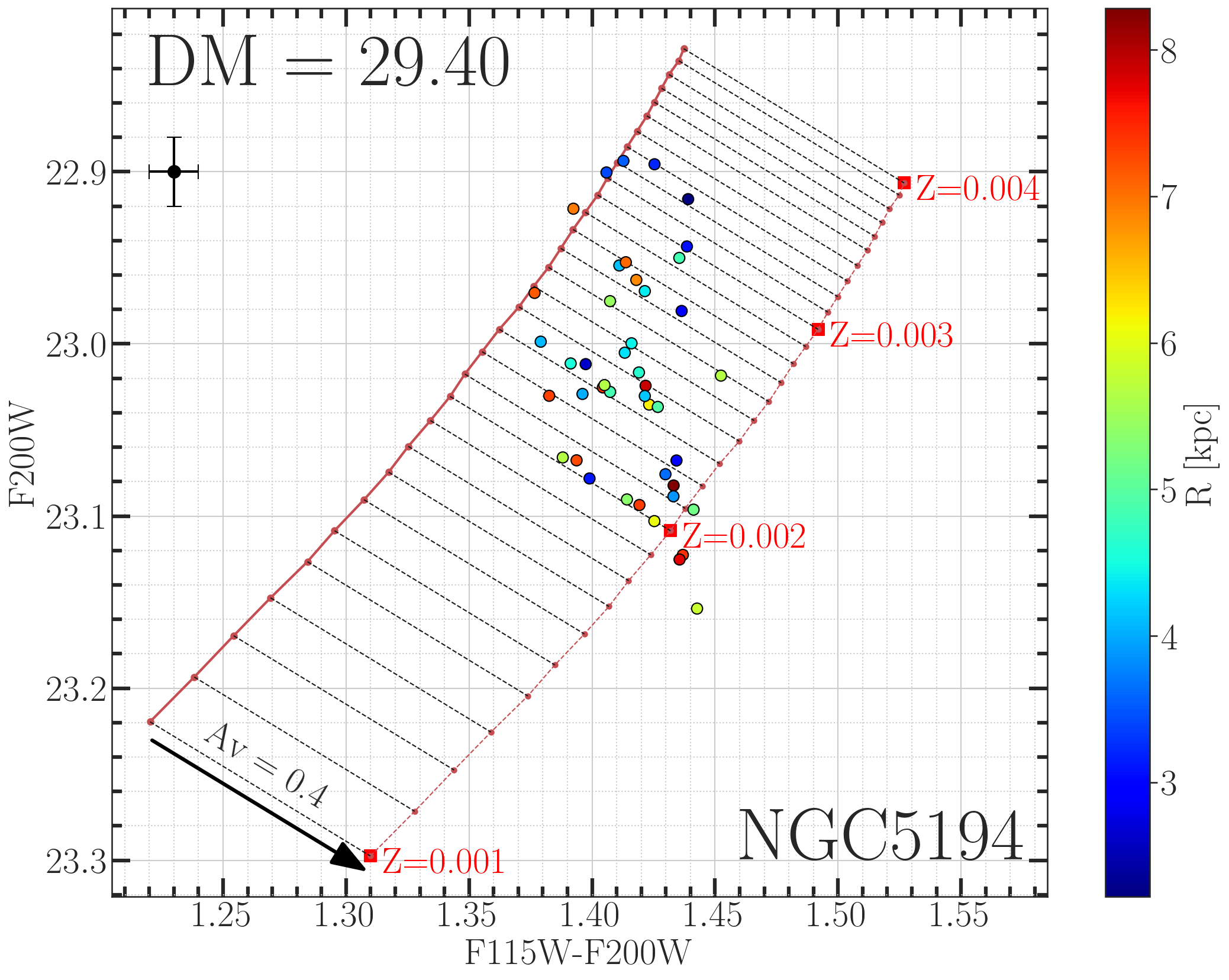}%
    \includegraphics[width=0.33\linewidth]{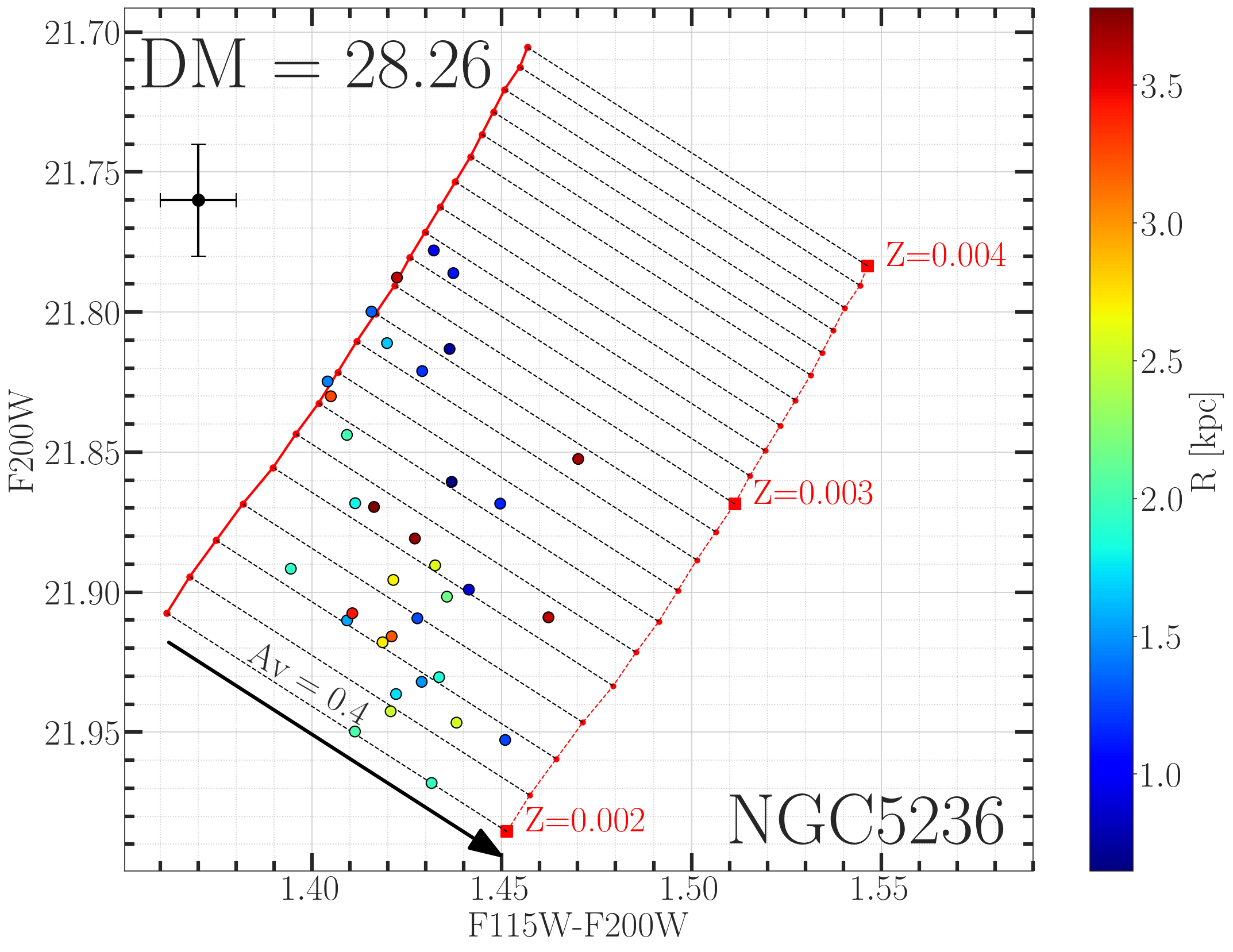}
    \includegraphics[width=0.33\linewidth]{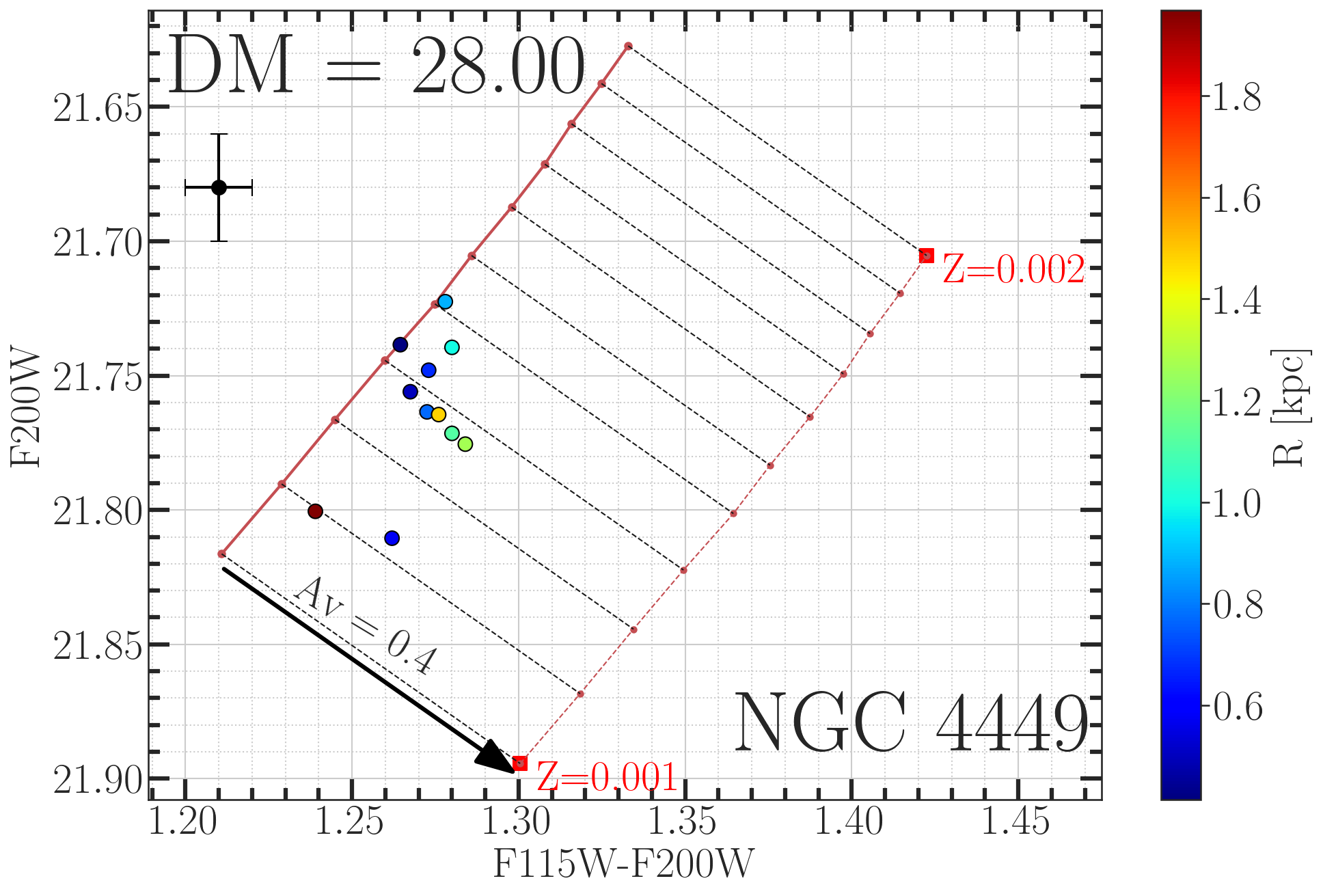}%
    \includegraphics[width=0.33\linewidth]{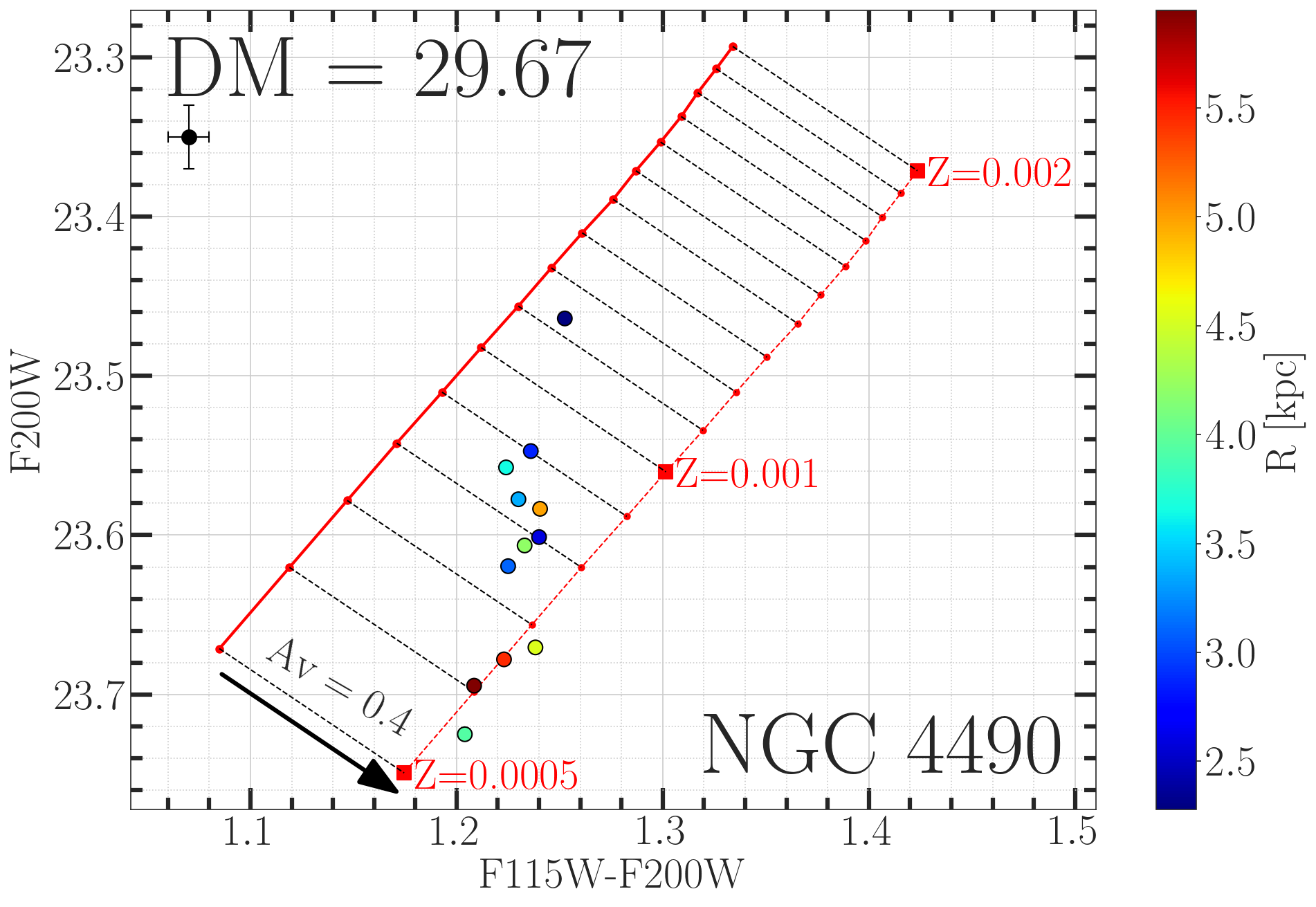}%
    \includegraphics[width=0.33\linewidth]{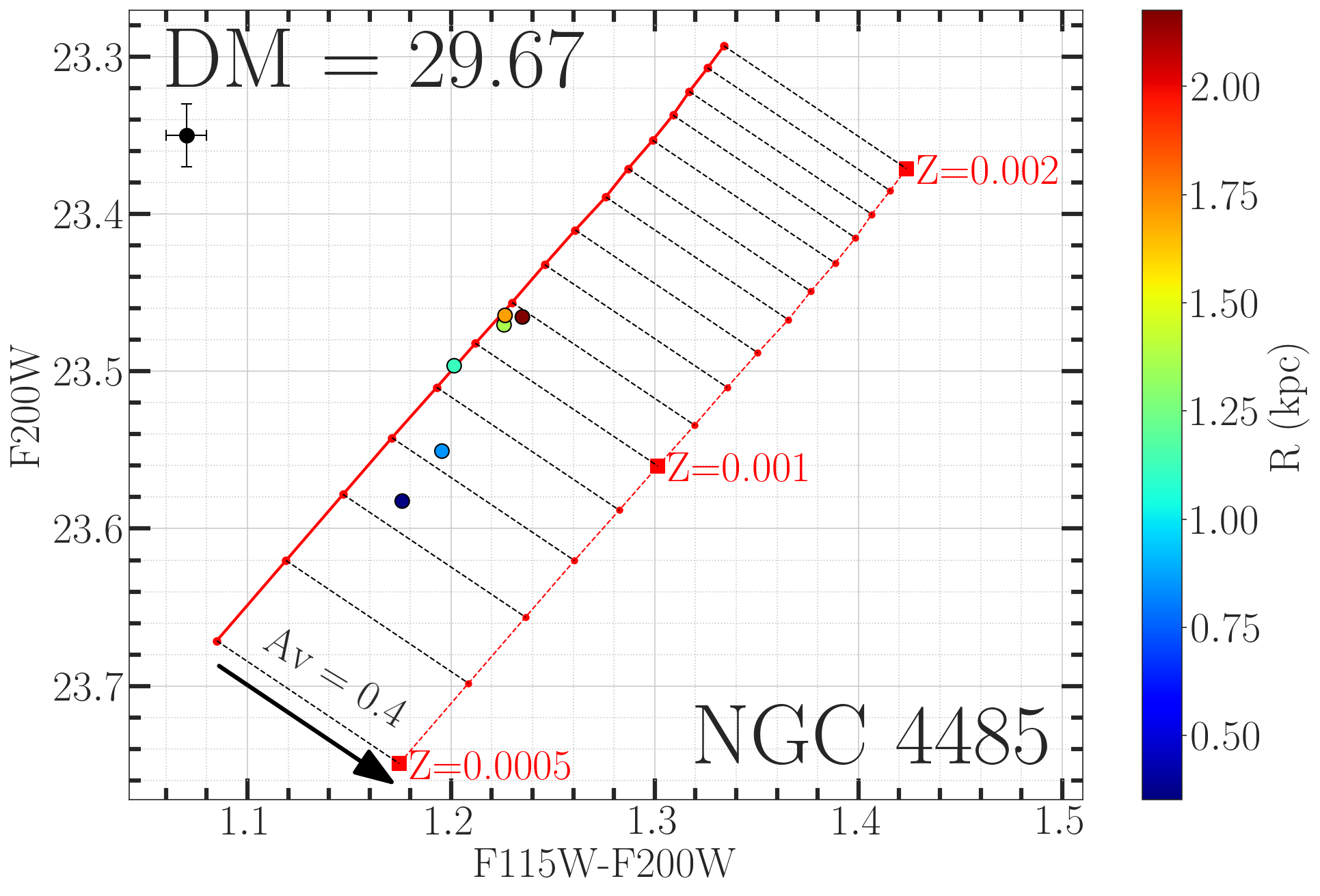}%
    \caption{TRGB color vs. TRGB magnitude for each spatial bin for the six galaxies. The red solid and dotted lines indicate the TRGB color and magnitude from PARSEC-COLIBRI models at different metallicities for the 13~Gyr population with $A_V$=0 and 0.4~mag, respectively. The separation of the grids along the metallicity curve corresponds to an incremental metallicity of 0.0001. The cross at the top-left corner indicates the typical error in TRGB color-magnitude measurement. The points are color-coded based on the galactocentric distance of the spatial bin that it represents, with the color-bar to the right showing the corresponding galactocentric distances. \textit{ Top:} (left to right) NGC\,628, NGC\,5194 and NGC\,5236. \textit{ Bottom:} (left to right) NGC\,4449, NGC\,4490 and NGC\,4485. In the top left panel the blue cross indicates the TRGB measurement of disk stars projected onto the Phantom Void.}
    \label{fig:dm}
\end{figure*}

\setcounter{footnote}{0}
In this work, we utilize the JWST-FEAST sample (PI: Angela Adamo, Program ID 1783), which comprises a set of nearby galaxies spanning a wide range of morphologies and stellar metallicities, namely NGC\,628 (M\,74)\footnote{This galaxy was studied in Paper I. All results pertaining to this galaxy utilized in this work are from Paper I}, NGC\,5194 (M\,51), NGC\,5236 (M\,83), NGC\,4449, and the interacting system NGC\,4490-4485. The basic properties of the galaxy sample, compiled from previous studies, are summarized in the first five columns of Table~\ref{tab:prop}. The JWST-FEAST NIRCam observations cover a field of view (FoV) of $6\arcmin \times 2\arcmin$, which encompasses a substantial part of the stellar disk in these galaxies. For the purposes of this study, we used observations in short-wavelength filters F115W and F200W, which have a spatial sampling of 0.031~arcsec\,pixel$^{-1}$. Data reduction and stellar photometry were carried out following the methods described in Paper~I, where we have also illustrated the utility of using F115W$-$F200W vs F200W color magnitude diagram (CMD) to obtain metallicities without the problems of metallicity-extinction degeneracy. The Hess diagrams of CMDs covering the entire field of view in each of the galaxies of the sample are shown in Fig.~\ref{fig:cmds}. The analysis of the CMDs to obtain the spatially-resolved metallicities is carried out using a spatial grid strategy that is described in the next section. We performed artificial star tests in each spatial grid for each galaxy to estimate the completeness and crowding uncertainties. The worst and best 50\% completeness magnitudes, corresponding to a grid in the inner and outer parts, respectively, are shown as gray curves for all the galaxies.

\section{TRGB color-magnitude measurement}

We adopted the edge-detection method in the F200W luminosity function (F200W-LF), referred to as the Sobel method \citep{trgb1},
to determine the TRGB magnitude and used the bi-weighted median to estimate the TRGB color as demonstrated in Paper I. The Sobel method involves two key parameters: the magnitude bin width ($\delta m$) and the magnitude smoothing scale ($\sigma_m$). Following the strategy outlined by \citet{Anand_2024}, we fix these parameters by selecting the values for which the resulting TRGB magnitude estimates are closest to the median. We found that $\delta m = 0.001$ and $\sigma_m = 0.05$ yield TRGB magnitudes closest to the median. Additionally, we find that the choice of binning and smoothing parameters introduces an uncertainty in the TRGB magnitude measurement of the order of $\pm 0.01$~mag. 

\subsection{Adaptive Spatial Binning}
\label{sec:adaptive_bin}
In order to maximize the utility of the JWST data, we carried out a spatially-resolved analysis of the CMD similar to the grid strategy adopted in Paper I. The sizes of the spatial grids were chosen such that the TRGB can be clearly identified in the F200W-LF for each grid. \citet{McQuinn_2019} demonstrated that for the TRGB to be sufficiently populated, the underlying stellar population must have a mass of at least $5\times10^7\,M_{\sun}$ using a Kroupa Initial Mass Function (IMF) \citep{kroupa_IMF}. For lower masses, the tip is not representative of the luminosity of the RGBs that are about to undergo helium flash, which would introduce errors additional to that intrinsic to the Sobel method. We made sure to fulfill this requirement with a view to get the maximum possible precision in the measured TRGB magnitude and colors. To estimate the minimum number of stars needed to meet this condition, we used TRILEGAL \citep{trilegal} to simulate the F115W$-$F200W versus F200W CMD of 4 and 13~Gyr stellar populations with a total mass of $5\times10^7$\,M$_{\sun}$ using a Kroupa IMF with mass range 0.1--350\,M$_{\sun}$ and a metallicity range of $Z$=0.001--0.006. We then counted the number of stars within a $0.1\times0.1$ color–magnitude window that covers 0.02 and 0.08 mag above and below the TRGB, respectively, accounting for photometric uncertainties. We find that $127\pm10$ and $58\pm6$  stars for 4 and 13~Gyr, respectively, within this TRGB color–magnitude window are sufficient to satisfy the stellar mass criterion. The chosen age values bracket the minimum and maximum mass of old populations, under the assumption that all fossil disks have stellar population at least as old as 4~Gyr.

We constructed adaptive spatial bins across the JWST-FEAST NIRCam FoV, such that each bin contains at least 150 stars within the TRGB color–magnitude window. This approach naturally yields higher spatial resolution in regions of higher stellar density, particularly toward the central regions of galaxies.  For the two giant spirals in the sample, NGC\,5194 and NGC\,5236, we utilized rectangular spatial bins similar to the strategy used for NGC\,628 in Paper~I. However, the relatively lower stellar densities in NGC\,4449, NGC\,4485, and NGC\,4490 required a different strategy. For these galaxies, we utilized concentric elliptical bins with the ellipses centered on the nucleus and aligned along the major axis with the ellipticity defined by the inclination angle of the galaxies. The areas of the spatial bins range between 0.5--2~kpc$^2$, with the lower values corresponding to denser inner parts. The number of spatial bins in each of our sample galaxies is listed in column~6 of Table~\ref{tab:prop}.

We generated CMDs for each spatial bin and measured the TRGB magnitudes and colors in all the six sample galaxies. The resulting TRGB color-magnitude measurements for all six galaxies after correcting for crowding are shown in Fig.~\ref{fig:dm}. The typical error bars in each measurement is shown at the top-left corner. The dispersion of the points is much larger than the measurement errors, and they are spread out along both the metallicity and extinction vectors. We also find a tendency of TRGBs being slightly bluer and brighter in the central regions as compared to the external parts, which can be inferred from the color-code used to represent each point. We discuss the implications of these tendencies in the next section.

\subsection{Distance modulus}

While the spread of points in Fig.~\ref{fig:dm} indicates spread in metallicity and extinction across the face of the galaxies, the quantitative determination of these physical quantities requires fixing the distances to each of these galaxies. 
The TRGB magnitude in the HST/F814W or the JWST/F090W provides the best standard candle for obtaining the distances to external galaxies \citep{McQuinn_2019}. The magnitudes in these filters are insensitive to metallicity over a wide range of possible metallicities of old stellar systems, which has allowed its calibration empirically within an uncertainty of $\sim$0.06~mag. All of our sample galaxies have previously measured TRGB distances which are listed in Table~\ref{tab:prop}. However, use of these distances led to several of our regions having TRGB measurements too blue and bright as compared to all possible isochrones. One such regions was the Phantom Void in NGC\,628 (Paper I, indicated by blue cross in the panel for NGC\,628 in Fig.~\ref{fig:dm}), the least extinction zone in the galaxy. This suggested that the distances measured using the halo TRGB F814W magnitudes need to be corrected for an internal extinction which could be different in different galaxies. We hence followed the same strategy as in Paper I of obtaining self-consistent distance moduli using the TRGB color-magnitude measurements of the least-extinction region(s) (bluest F115W$-$F200W points in Fig.~\ref{fig:dm}) and comparing it with the PARSEC-COLIBRI\footnote{We utilized PARSEC 1.2S + COLIBRI S\_37 + COLIBRI S\_35 + COLIBRI PR16, generated using web interface \href{https://stev.oapd.inaf.it/cgi-bin/cmd\_3.9}{CMD 3.9} with  n$_{\rm inTPC}$=10 and $\eta_{\rm Reimers}$=0.2} isochrones \citep{parsec, parsec2, parsec_colibri}. The resulting distance moduli are given in column~7 of Table~\ref{tab:prop}. We note that within the systematics of TRGB-based distance measurements (errors of 0.06~mag in our measurements and more than 0.1~mag in the HST F814W-based measurements), our values are in agreement with previous estimates.

\begin{figure*}
    \centering
    \includegraphics[width=0.5\linewidth]{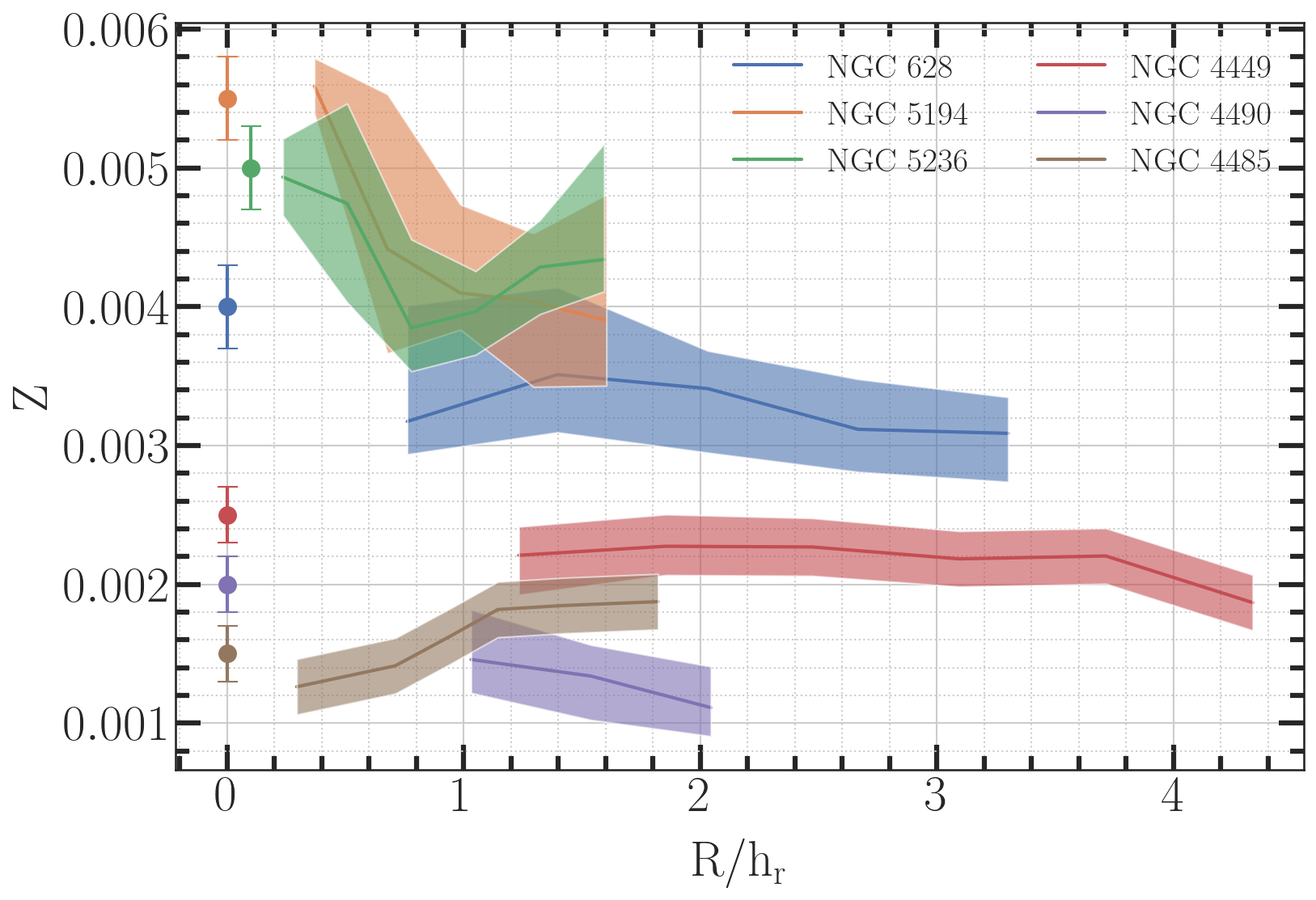}%
    \includegraphics[width=0.5\linewidth]{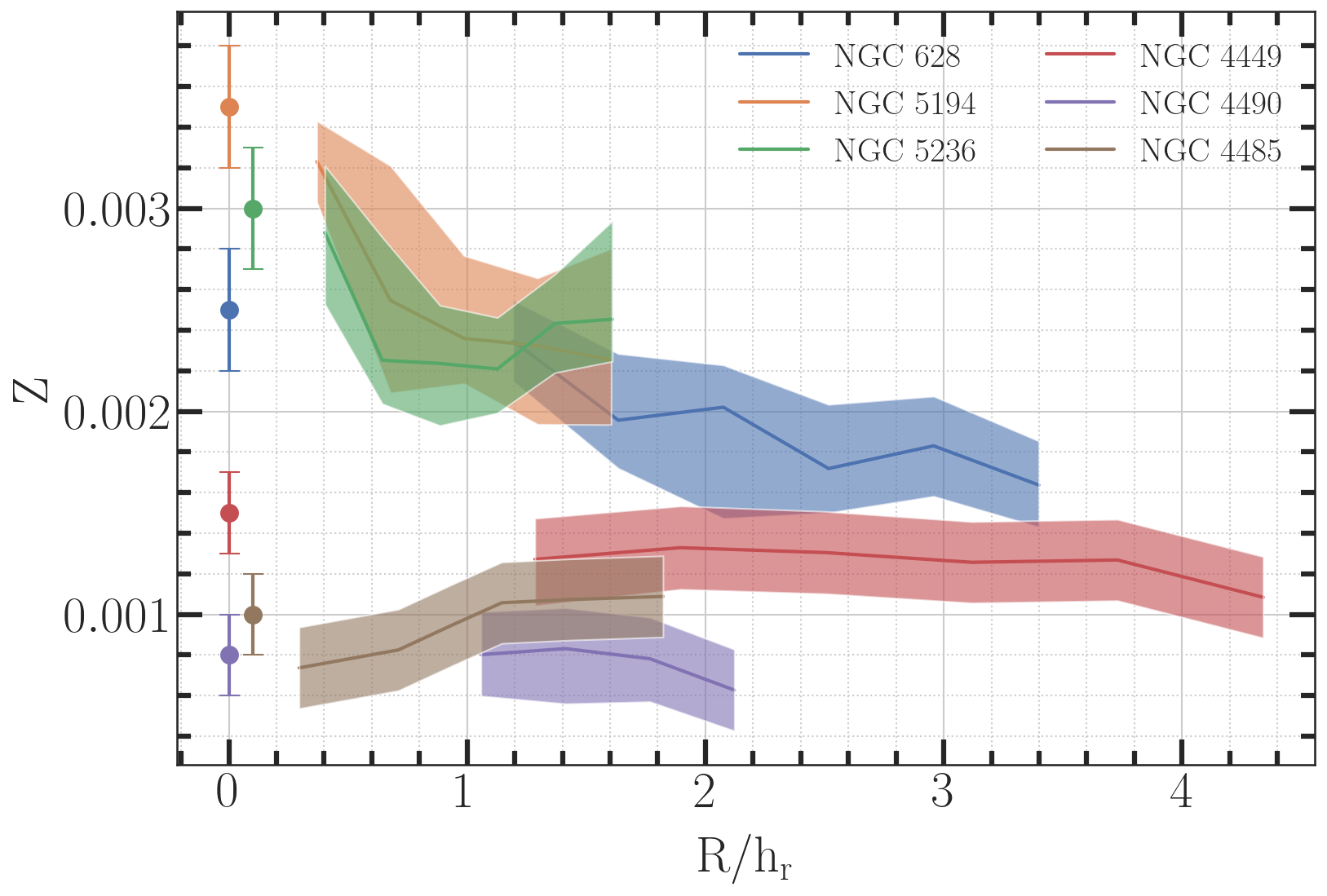}
    \caption{Stellar metallicity vs. Galactocentric radii normalized with scale radius. The width of the colored bands indicates dispersion in metallicity among regions in the same radial bin. Typical error bar on each metallicity measurement is shown by a vertical line at the inner-most radial bin for each galaxy. \textit{Left:} 4~Gyr. \textit{Right:} 13~Gyr.}
    \label{fig:met_grad}
\end{figure*}

\subsection{Stellar Mass of the RGB population}

For a given IMF and a population of fixed age, the number of TRGBs in a fixed color-magnitude window  scales with the mass of the stellar population. Hence, we can obtain the stellar mass in each spatial grid by simple star counts in the TRGB window. We used the stellar mass and area of each spatial bin to obtain the stellar mass surface density as a function of galactocentric radius for each galaxy, following the method of \citet{Harmsen2017} and \citet{Bell2017} for halo stellar masses. In galaxy disks, however, the TRGB window may contain RGB stars younger than 4\,Gyr. To correct for this contamination, we counted stars in a $0.1\times0.1$\,mag color-magnitude window located 0.5~mag above the TRGB in the asymptotic giant branch (AGB), representing stars younger than 4~Gyr. We then used the AGB-to-RGB ratio from TRILEGAL simulations assuming a constant star formation rate from 13~Gyr to 100~Myr to estimate and remove the contribution of RGB stars younger than 4~Gyr from the TRGB window, yielding the corrected mass surface density profile of each galaxy. The resulting profiles are well described by an exponential, characteristic of galaxy disks. We fitted each profile with an exponential function and integrated it from zero to infinity to obtain the total stellar mass, at birth, of the population that produced the RGB stars. Due to JWST/NIRCam coverage and crowding in the central region, the fraction of mass obtained from extrapolation varies widely between samples; the fraction of mass directly observed is indicated in column~14 of Table~\ref{tab:prop}. Because stars at the RGB tip undergo helium flash at ages between 4 and 13~Gyr, we calculated stellar masses for these two extreme ages, which give the lower and upper mass limits listed in columns~11 and 12 of Table~\ref{tab:prop}. The derived masses and best-fit scale lengths carry important information about the formation and growth of the early disk, but interpreting them requires a detailed star formation history (SFH) analysis, which will be presented in a future paper (Ck et al., in preparation).

\section{Results and Discussion}

\subsection{Stellar Metallicity Gradients}

After fixing the distance modulus, each point in the TRGB color-magnitude diagram (Fig.~\ref{fig:dm}) is translated along the extinction vector until it intersects the locus defined by the TRGB color-magnitude relation of PARSEC-COLIBRI isochrones spanning different metallicities. The displacement along the extinction vector yields the extinction, while the point of intersection provides the corresponding stellar metallicity. As illustrated in Paper~I, the TRGB point is $\sim$7 times more sensitive to metallicity variations than to age variations between 4 and 13~Gyr. We however take into account this small sensitivity to ages by calculating  independent metallicities at 4 and 13~Gyr. 

In Fig.~\ref{fig:met_grad}, we show the stellar metallicity as a function of galactocentric radius for all the galaxies in this work, with the left and right panels showing the results for 4~Gyr and 13~Gyr RGBs, respectively. The metallicity gradient plots at these two assumed RGB ages are similar with the derived metallicity systematically higher by $\sim$0.25~dex for the younger age. The metallicity gradient of the fossil disk is negative in the spirals of our sample (NGC\,628, NGC\,5194, NGC\,5236 and NGC\,4490) and flat for the irregular galaxy NGC\,4449, consistent with trends from nebular abundance analyses where available \citep{ngc628_Z_HII, m51_Z_HII, m83_Z_HII, ngc4449_Z_HII, ngc4490_Z_HII}. The interacting companion NGC\,4485, by contrast, shows a positive gradient.

\begin{figure*}
    \centering
    \includegraphics[width=0.5\linewidth]{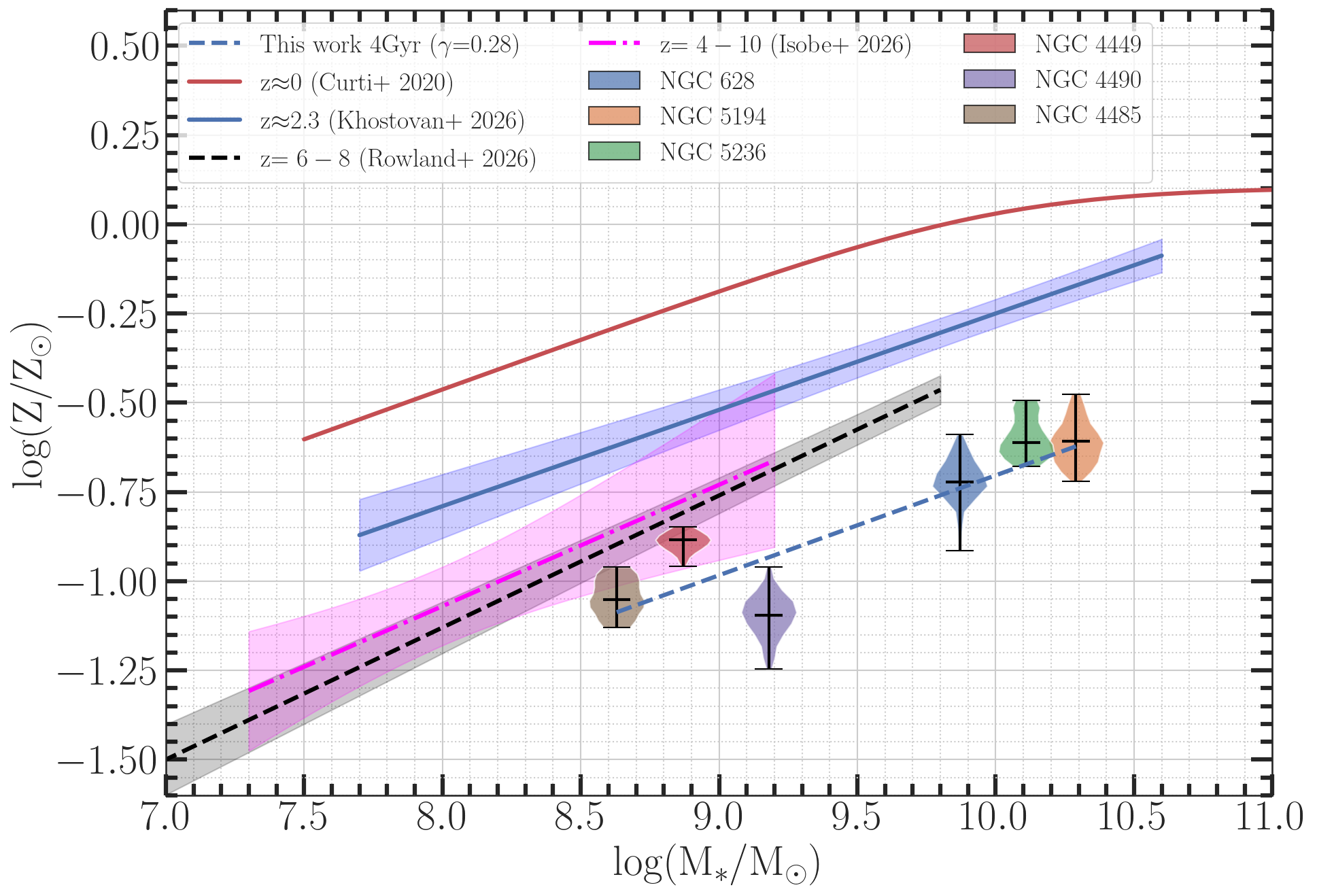}%
    \includegraphics[width=0.5\linewidth]{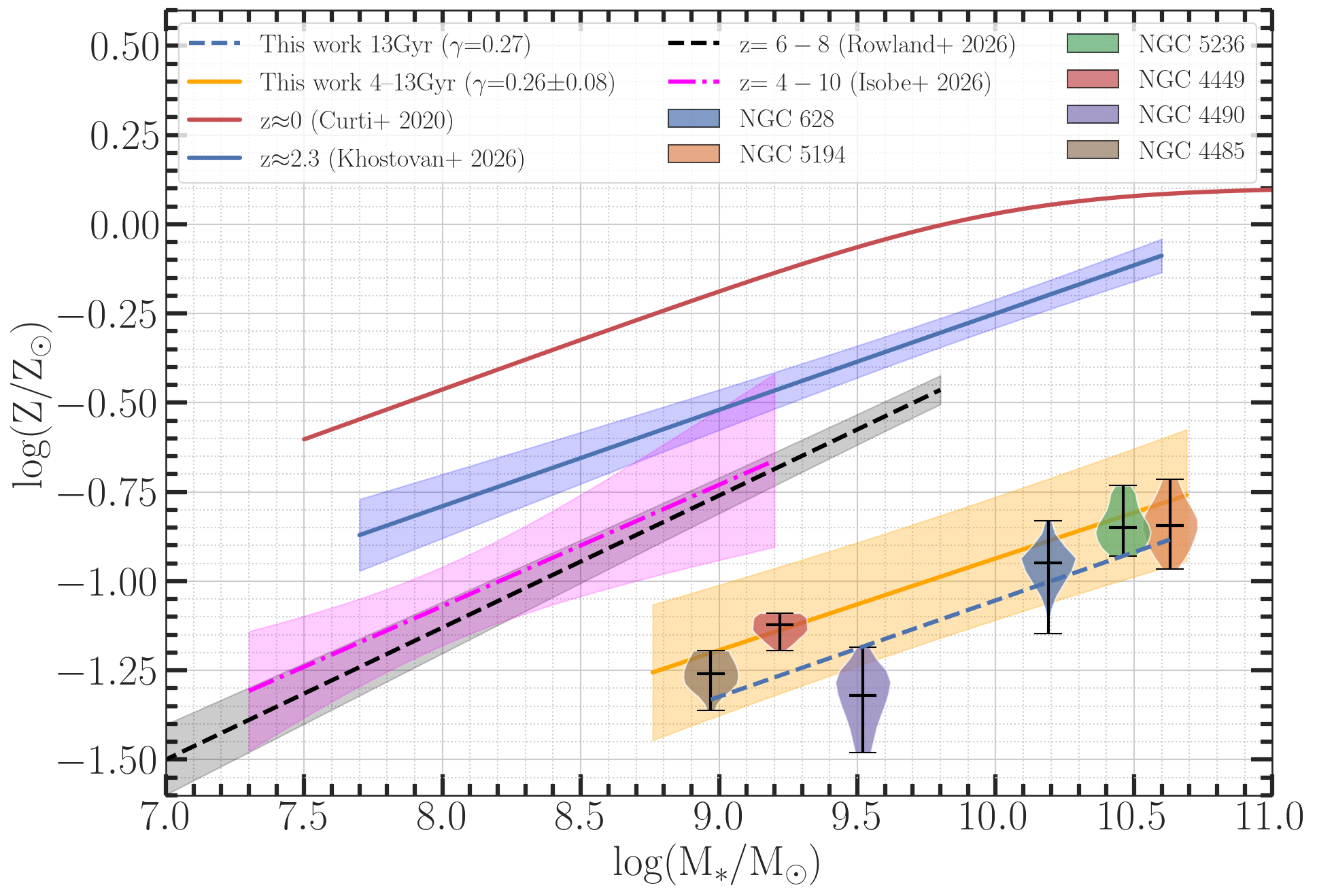}
    \caption{Mass-metallicity relation derived from TRGB color-magnitude measurements assuming a stellar population age of.(\textit{left}) 4~Gyr. (\textit{right}) 13~Gyr. The orange line shows fitted line after randomly choosing age of RGB population for each galaxy. The solid red line shows the MZR at $z\approx0$ from \citet{curti}, while the solid blue line shows the MZR at $z\approx2.3$ from \citet{aurora_survey}. The black dashed line represents the MZR at $z=6$--8 from \citet{rebels}, and the magenta dashed line represents the MZR at $z=4$--10 obtained from \citet{jades}. The dashed blue and red lines show the MZR derived in this work using the stellar mass density weighted median and maximum metallicities, respectively, for each galaxy. }

    \label{fig:mass_z}
\end{figure*}

\subsection{Mass Metallicity Relation}

In Fig.~\ref{fig:mass_z}, we present the MZR of the fossil stellar populations, constructed from the TRGB-based stellar metallicities as a function of the stellar mass of the RGB population for assumed population ages of 4~Gyr (left) and 13~Gyr (right), where the metallicities are normalized to the solar metallicity, $Z_{\sun}=0.017$. Given the metallicity gradient discussed above, each galaxy shows a range of metallicities, represented by violin plots showing the stellar-density-weighted metallicity distribution functions. The plots show a clear positive correlation between the RGB-derived mass and metallicities. The plots for the two ages are similar with the 13~Gyr old MZR shifted by 0.34~dex and 0.25~dex along the mass and metallicity axes, respectively, with respect to the 4~Gyr relation. The RGB population of each galaxy may have any age between 4 and 13\,Gyr, and the derived stellar mass and metallicity both depend on the adopted age. To propagate this uncertainty, we generated $N=50{,}000$ Monte Carlo realizations, drawing an RGB age for each galaxy from a uniform distribution over this range.
In each realization, we fit a power law to the resulting mass--metallicity relation. The median slope is $0.26\pm0.08$ (Fig.~\ref{fig:mass_z} (right)), where the uncertainty is the standard deviation of the slope distribution along with the rms of the fit.

For comparison, we also show MZRs derived using different observational techniques over a range of redshifts covering local galaxies ($z=0$), intermediate redshifts ($z=2.3$) and galaxies at redshifts as high as $z=8$, obtained using recent JWST observations. These redshifts correspond to lookback times  0, 10.8 and 13.1~Gyr, respectively, for the cosmology with H$_0$=69.6 km/s/Mpc, $\Omega_M=$0.286, $\Omega_\Lambda$=0.714. Since the comparison galaxies are observed at high redshift, their stellar populations formed at cosmological epochs similar to those of the oldest populations in the fossil disks of local galaxies. Their SED-based masses \citep[e.g.][]{bc03} are also comparable to our birth masses, since for such young populations the surviving and birth masses differ only slightly. The metallicities are derived with nebular techniques \citep{Peimbert_2017} that trace the ISM metallicity. On the other hand, the RGB metallicities are representative of the ISM at the formation epoch of these stars, which corresponds to lookback times greater than 4~Gyr ($z>0.36$). This implies that the MZR for the assumed RGB age of 13~Gyr should be comparable to the MZR of galaxies of $z\approx8$.

For a galaxy of given stellar mass, the gas metallicities increase from past to present, with the bulk of the increase occurring during the cosmic noon \citep{liu2026}. The observed RGB metallicities are in the range of gas metallicities of galaxies at $z>8$, suggesting that the formation epochs of the RGB stars are closer to 13~Gyr than 4~Gyr. However the 13~Gyr MZR is shifted to the right along the mass axis by an order of magnitude. This shift is indicating that our RGB-derived masses include a contribution from metal-poor populations that are accreted at later epochs (ex-situ stars) as expected in the hierarchical galaxy formation scenarios \citep{Yetli2026}. It is likely that the fossil stellar disk was in place 13~Gyr ago in massive galaxies, given that even low-mass systems with current star formation such as Leo P have an underlying 13~Gyr population \citep{leop}. Recent JWST observations have found disk galaxies at $z\sim6$, suggesting disks are formed very early \citep{Ferreira_2023, Robertson_2023}. Hydrodynamical simulations also predict formation of disk galaxies at these high redshifts \citep{Feng_2015,Semenov_2025}. 

\section{Conclusions}

In this Letter, we analyzed the RGB stars in the JWST images of six nearby galaxies to construct the MZR of the fossil disks. The availability in recent years of the nebular abundances for high redshift galaxies, which probe the metallicities at the formation epoch of the fossil stars, allows a direct comparison of the MZR during the early epochs of galaxy formation using two complementary approaches. We find that the fossil disks follow a power-law MZR with an index $\gamma\approx0.26$, a value similar to that found across redshifts. The metallicity of the fossil stars compares well with the metallicity prevalent at the time of the formation of these stars as probed by the $z>8$ galaxies. The current mass of the fossil stellar disks, however, is an order of magnitude higher, suggesting incorporation of large quantities of metal-poor stars through accretion over the lifetime of galaxies. This study illustrates the potential of JWST/NIRCam images of nearby galaxies in complementing the JWST study of high redshift emission-line galaxies. In a future work, we would extend the MZR relation to lower metallicities using the recently available NIRCam images of a limited number of low-mass galaxies.

\begin{acknowledgments}
ACk expresses gratitude to the Secretaría de Ciencia, Humanidades, Tecnología e Innovación (SECIHTI) for funding his PhD through a grant. AB acknowledges the support received through the Chinese Academy of Sciences (CAS) President’s International Fellowship Initiative (PIFI 2025). LG and AB acknowledge support from an INAF Large Grant 2024. SM's work has received support from France 2030 through the project named Académie Spatiale d'I\^{l}e-de-France \url{https://academiespatiale.fr/} managed by the National Research Agency under bearing the reference ANR-23-CMAS-0041. We also thank Dr. Raul Naranjo and Dr. Manuel Zamora for providing access to the Mixli cluster at INAOE, which was crucial for the computational work carried out in this research.

Computational resources were also provided by El Laboratorio Nacional de Supercómputo del Sureste de México (LNS), under project numbers \textsf{202404072C} and \textsf{202404078C}.

Claude (Anthropic; Claude Sonnet 5.5) was used during manuscript preparation
for language editing and structural suggestions. The authors reviewed and
verified all text and take full responsibility for the content of this paper.
\end{acknowledgments}





%

\section*{Data Availability}
This work is based on data obtained with the NASA/ESA/CSA James Webb Space Telescope, archived at the Mikulski Archive for Space Telescopes at the Space Telescope Science Institute. The institute is managed by the Association of Universities for Research in Astronomy, Inc., under NASA contract NAS 5-03127 for JWST. These data were collected as part of program \#1783 (JWST-FEAST; PI: Angela Adamo) and can be accessed at doi:\href{https://doi.org/10.17909/6hpm-nr53}{10.17909/6hpm-nr53}.

\facilities{JWST/NIRCam.}

\software{ \href{http://americano.dolphinsim.com/dolphot/nircam.html}{DOLPHOT} \citep{dolphot_1, dolphot_2}, TRILEGAL \citep{trilegal}, Astropy \citep{astropy}, NumPy \citep{numpy}, and Matplotlib \citep{matplotlib}}.


\bibliography{references}{}
\bibliographystyle{aasjournalv7}



\end{document}